\documentclass[%
superscriptaddress,
showpacs,
nofootinbib,
amsmath,amssymb,
prc,
twocolumn,
floatfix ]%
{revtex4-1}

\usepackage{color}
\usepackage{CJK}

\usepackage{graphicx}% Include figure files
\usepackage{dcolumn}% Align table columns on decimal point
\usepackage{bm}% bold math
\usepackage{hyperref}% add hypertext capabilities
\allowdisplaybreaks

\begin{document}

\begin{CJK*}{UTF8}{}
\title{Extended time-dependent generator coordinate method study of induced fission (II): total kinetic energy distribution}
\CJKfamily{gbsn}
\author{Jie Zhao (赵杰)}%
%\email{zhaojie@alumni.itp.ac.cn}
\affiliation{Center for Circuits and Systems, Peng Cheng Laboratory, Shenzhen 518055, China}
%\affiliation{Center for Quantum Computing, Peng Cheng Laboratory, Shenzhen 518055, China}
%\affiliation{Microsystem and Terahertz Research Center and Insititute of Electronic Engineering, 
%	China Academy of Engineering Physics, Chengdu 610200, Sichuan, China}
% \affiliation{Physics Department, Faculty of Science, University of Zagreb, Bijeni\v{c}ka Cesta 32,
%             Zagreb 10000, Croatia}
% \affiliation{CAS Key Laboratory of Theoretical Physics,
%              Institute of Theoretical Physics, Chinese Academy of Sciences, Beijing 100190, China}
\author{Tamara Nik\v{s}i\'c}%
%\email{tniksic@phy.hr}
\affiliation{Physics Department, Faculty of Science, University of Zagreb, Bijeni\v{c}ka Cesta 32,
        	      Zagreb 10000, Croatia}             
\author{Dario Vretenar}%
%\email{vretenar@phy.hr}
\affiliation{Physics Department, Faculty of Science, University of Zagreb, Bijeni\v{c}ka Cesta 32,
              Zagreb 10000, Croatia}
 \affiliation{ State Key Laboratory of Nuclear Physics and Technology, School of Physics, Peking University, Beijing 100871, China}            
%\CJKfamily{gbsn}              
%\author{Shan-Gui Zhou (周善贵)}%
%\email{sgzhou@itp.ac.cn}
 %\affiliation{CAS Key Laboratory of Theoretical Physics, Institute of Theoretical Physics, Chinese Academy of Sciences, Beijing 100190, China}
 %\affiliation{School of Physical Sciences, University of Chinese Academy of Sciences, Beijing 100049, China}
 %\affiliation{Center of Theoretical Nuclear Physics, National Laboratory of Heavy Ion Accelerator, Lanzhou 730000, China}
 %\affiliation{Synergetic Innovation Center for Quantum Effects and Application, Hunan Normal University, Changsha 410081, China}

\date{\today}

\begin{abstract}
A model is developed to calculate the total kinetic energy (TKE) distribution of fission fragments in the framework of the time-dependent generator coordinate method (TDGCM), extended to include dissipation effects in the description of induced fission dynamics. Starting from an expression for the dissipative term in the GCM Hamiltonian that determines the time evolution of a statistical collective wave function, derived in the first part of this work, the integrated flux of the probability current through the scission hyper-surface can be computed at different temperatures. The kinetic energy at scission for a specific pair of fragments at a given temperature is determined by the energy balance formula. By folding the kinetic energies of the fragments with the flux of the probability current, the TKE distribution is calculated. The method is illustrated with a calculation of the TKE distribution for induced fission of $^{228}$Th, in the 3D space of collective coordinates: axially-symmetric quadrupole and octupole deformations, and nuclear temperature. 
\end{abstract}

\maketitle

\end{CJK*}

\bigskip

\section{Introduction~\label{sec:Introduction}}
The time-dependent generator coordinate method (TDGCM) \cite{krappe12,schunck16,younes19,bender20,verriere20} has been successfully applied in theoretical studies of various aspects of the nuclear fission process 
\cite{Regnier2016_CPC200-350,Regnier2018_CPC225-180,Regnier16,Regnier19,Tao2017_PRC96-024319,Zhao2019_PRC99-014618,Zhao2019_PRC99-054613,Zhao2020_PRC101-064605,Zhao2021_PRC104-044612,Verriere21,schunck22,Li2022_PRC106-024307}. The GCM presents a fully microscopic, quantum mechanical approach to nuclear dynamics, in which the nuclear wave function is determined by a superposition of generator states that are functions of collective coordinates, such as shape variables and pairing degrees of freedom. In the Gaussian overlap approximation (GOA) of the TDGCM, in particular, a time-dependent Schr\"odinger equation governs the evolution of the nuclear wave function in the space of collective coordinates. The TDGCM + GOA can be applied to an adiabatic description of the entire fission process, from the quasi-stationary initial state to the outer fission barrier (saddle point), and to the scission of the nucleus into fission fragments. In many cases the calculated fission yields (charge and mass distribution of fragments) are in quantitative agreement with data, especially when the framework is extended to finite temperature. However, since only collective degrees of freedom are explicitly taken into account, the standard TDGCM does not include any dissipation mechanism and, therefore, fails to describe the strongly dissipative dynamics in the saddle-to-scission phase of the fission process. In particular, in the absence of dissipation, all the potential energy difference between saddle and scission is converted into collective kinetic energy during the saddle-to-scission evolution. Even though the calculated kinetic energies of the fragments qualitatively reproduce the empirical charge or mass dependence, in general they are systematically too large when compared to data \cite{Ren2022_PRC105-044313}. 

In the first part of this work \cite{Zhao2022_PRC105-054604}, we have extended TDGCM to allow for dissipation effects in the description of induced fission dynamics. The extension is based on the quantum theory of dissipation for nuclear collective motion, introduced by Kerman and Koonin in Ref.~\cite{Kerman1974_PS10-118}. The GCM generating functions are generalized to include not only self-consistent deformation-constrained mean-field product states of lowest energy, but also excited states. The resulting equation of motion in the collective coordinates, that is, the Hamiltonian and overlap kernels, explicitly depend on the excitation energy. With the  assumption of a narrow Hamiltonian kernel, an expansion in a power series in collective momenta leads to a Schr\"odinger-like equation that explicitly includes a dissipation term, proportional to the momentum of the statistical wave function. By expressing the excitation energy in terms of nuclear temperature, the model can be formulated as an  extended temperature-dependent TDGCM, in which the Helmholtz free energy plays the role of the collective potential, and the collective inertia is calculated in the finite-temperature perturbative cranking approximation. The dissipation-extended TDGCM  has been applied in an illustrative calculation of fission yields of $^{228}$Th, and the results have been compared to available data for photo-induced fission of $^{228}$Th \cite{Schmidt2000_NPA665-221}. 

In the present study we further develop a consistent procedure, based on the model introduced in the first part of this work, to calculate the total kinetic energy (TKE) distribution of fission fragments. The method is developed in Sec.~\ref{sec:model}. Section~\ref{sec:results} contains an illustrative calculation of the TKE distribution for induced fission of $^{228}$Th, in comparison with data for photo-induced fission. In Sec.~\ref{sec:summary} we summarize the results and present a brief outlook for future studies.

%%%%%%%%%%%%%%%%%%%%%%%%%%%%%%%%%%%%%%%%%%%%
\section{\label{sec:model}Theoretical framework}
%%%%%%%%%%%%%%%%%%%%%%%%%%%%%%%%%%%%%%%%%%%%
In Ref.~\cite{Zhao2022_PRC105-054604} we have extended the time-dependent generator coordinate method (TDGCM) description of induced fission dynamics, by including an energy dissipation term derived from the quantum theory of dissipation introduced in Ref.~\cite{Kerman1974_PS10-118}. This extension follows from a generalization of the GCM generating functions that includes excited intrinsic states. With the assumption of a narrow hamiltonian kernel, a   
time-dependent Schr{\"o}dinger-like equation for the statistical collective wave function is obtained
\begin{equation}
\label{eq:tdgcmgoa_ft}
\begin{aligned}
i\hbar \partial_t \psi (\bm{q}, T; t)
	=& \left[ V(\bm{q}, T)
		+ \bm{P} \frac{1}{2 \mathcal{M}(\bm{q}, T)} \bm{P} \right] \psi (\bm{q}, T; t) \\
		&+ \frac{i}{2} \int \left\{ \bm{P}, \bm{\mathcal{O}}(\bm{q}; T, T^{\prime}) \right\} \psi (\bm{q}, T^{\prime}; t) dT^{\prime},
\end{aligned}
\end{equation}
where $\bm{q}$ denotes the set of collective coordinates such as, for instance, the quadrupole and octupole deformation parameters, $T$ is the nuclear temperature, 
$\bm{P} = -i\hbar(\partial / \partial \bm{q})$ is the collective momentum,
$V(\bm{q}, T)$ and $\mathcal{M}(\bm{q}, T)$ are the collective potential and mass tensor, respectively, 
$\bm{\mathcal{O}}(\bm{q}; T, T^{\prime}) = \bm{\mathcal{\eta}}(\bm{q}; T, T^{\prime}) d\epsilon(T) / dT$, $\epsilon(T)$ is the excitation energy as function of temperature, 
and $\bm{\eta}(\bm{q}; T, T^{\prime})$ is the dissipation function. Therefore, in addition to the collective potential and kinetic energy terms, equation (\ref{eq:tdgcmgoa_ft}) explicitly includes a dissipation term proportional to the momentum of the statistical wave function. For details of the derivation, we refer the reader to Ref.~\cite{Zhao2022_PRC105-054604}. 

In the present study we employ the self-consistent multidimensionally-constrained (MDC) relativistic Hartree-Bogoliubov (RHB) 
model ~\cite{Lu2014_PRC89-014323,Zhao2017_PRC95-014320} at finite
temperature~\cite{Goodman1981_NPA352-30,Egido1986_NPA451-77,Zhao2019_PRC99-014618}, to calculate the microscopic single-nucleon wave functions and occupation probabilities that determine the collective potential, mass tensor and entropy. 
The particle-hole channel is specified by the choice of the relativistic energy density functional DD-PC1~\cite{Niksic2008_PRC78-034318}, 
while pairing correlations are taken into account in the Bardeen-Cooper-Schrieffer (BCS) approximation with a finite-range separable 
pairing force~\cite{Tian2009_PLB676-44}. 
The parameters of the pairing interaction have been adjusted to reproduce the empirical pairing gaps 
in the mass region considered in this study~\cite{Zhao2019_PRC99-054613}. 
The nuclear shape is parameterized by the deformation parameters
\begin{equation}
 \beta_{\lambda\mu} = {4\pi \over 3AR^\lambda} \langle Q_{\lambda\mu} \rangle.
\end{equation}
The shape is assumed to be invariant under the exchange of the $x$ and $y$ axes, 
and all deformation parameters $\beta_{\lambda\mu}$ with even $\mu$ can be included simultaneously.
The self-consistent relativistic mean-field (RMF+BCS) equations are solved by an expansion in the 
axially deformed harmonic oscillator (ADHO) basis~\cite{Gambhir1990_APNY198-132}.
In the present study calculations have been performed 
in an ADHO basis truncated to $N_f = 20$ oscillator shells.
The thermodynamical potential relevant for deformation effects is the Helmholtz free energy $F(T) = E(T) - TS$, evaluated at constant temperature $T$, 
where $E(T)$ is the binding energy of the deformed nucleus, and the deformation-dependent energy landscape is obtained in a self-consistent
finite-temperature mean-field calculation with constraints on the mass multipole moments $Q_{\lambda\mu} = r^{\lambda} Y_{\lambda\mu}$.

In the present analysis the collective coordinates $\bm{q}$ correspond to the quadrupole $\langle Q_{20} \rangle$ and 
octupole  $\langle Q_{30} \rangle$ mass multipole moments.
The collective potential is, therefore, $V(\bm{q}, T) = \epsilon(T) + F(\bm{q}, T)$, where $F(\bm{q}, T)$ is the Helmholtz free energy
normalized to the corresponding value at the equilibrium RMF+BCS minimum at temperature $T$. 
The internal excitation energy $\epsilon(T)$ of a nucleus at temperature $T$ is defined as the difference between the total binding energy 
of the equilibrium RMF+BCS minimum at temperature $T$ and at $T=0$.
The mass tensor $\mathcal{M}(\bm{q}, T)$ is calculated 
in the finite-temperature perturbative cranking approximation~\cite{Zhu2016_PRC94-024329,Martin2009_IJMPE18-861}.

Equation (\ref{eq:tdgcmgoa_ft}) describes nuclear collective motion with dissipation. 
In addition to the non-dissipative potential and kinetic energy terms, 
the dissipative channel coupling is proportional to the momentum of the collective wave function. 
As explained in Ref.~\cite{Kerman1974_PS10-118,Zhao2022_PRC105-054604}, 
we choose the dissipation function $\bm{\mathcal{\eta}}(\bm{q}; T, T^{\prime})$ to be of the form
\begin{equation}
\label{eq:eta}
 \bm{\mathcal{\eta}}(\bm{q}; T, T^{\prime}) = 
 \begin{cases}
	 0 			             & \beta_{2}< \beta_{2}^{0}         \\
	 \bm{\eta}(T, T^\prime)  & \beta_{2} \geq \beta_{2}^{0},
 \end{cases} 
\end{equation}
where the matrix elements $\bm{\eta}(T, T^\prime)$ are Gaussian random variables. $\beta_{2}^{0}$ is set to $1.5$, 
which is slightly beyond the second fission barrier for the example of fission of $^{228}$Th, that will be considered in the next section. 
The root-mean-square value of the Gaussian distribution of the $\bm{\eta}(T, T^\prime)$ random variables reads  $\displaystyle \gamma \sqrt{ \log[\rho(T)] \log[\rho(T^{\prime})] }$. 
In this expression $\rho(T)$ is the intrinsic nuclear level density calculated at the RMF+BCS equilibrium minimum at temperature T, while $\gamma$ is an adjustable parameter. 
For further details we refer the reader to Refs.~\cite{Kerman1974_PS10-118,Zhao2022_PRC105-054604}.

 To model the dynamics of the fission process we follow 
 the time-evolution of an initial wave packet $\psi(\bm{q},T,t=0)$, built 
as a Gaussian superposition of quasi-bound states $g_k$
\begin{equation}
\label{eq:initial-wave-packet}
\psi(\bm{q},T,t=0) = \sum_k{e^{(E_k-\bar{E})^2/\left(2\sigma^2\right)}g_k(\bm{q},T)},
\end{equation}
where the value of the parameter $\sigma$ is set to 0.5 MeV. The collective states $\{g_k(\bm{q},T)\}$ 
and the corresponding energies $E_k$ 
are solutions of the stationary eigenvalue equation, in which the original collective potential
is replaced by a new potential $V^\prime(\bm{q},T)$ that is obtained by extrapolating the inner potential
barrier with a quadratic form. A more detailed description of this procedure can be found in Ref.~\cite{Regnier2018_CPC225-180}. 
The average energy of the collective initial state is calculated as
\begin{equation}
\label{eq:ecoll}
E^{*}_{\rm{coll}} = \left\langle \psi(\bm{q},T,t=0) \right| H_{\rm coll} \left| \psi(\bm{q},T,t=0)\right\rangle,
\end{equation}
where $H_{\rm coll}=V(\bm{q}, T) + \bm{P}  [2 \mathcal{M}(\bm{q}, T)]^{-1} \bm{P}$,
and the mean energy $\bar{E}$ in Eq.~(\ref{eq:initial-wave-packet}) is adjusted iteratively to obtain the chosen value of $E^{*}_{\rm{coll}}$.

The time-evolution is described by Eq.~(\ref{eq:tdgcmgoa_ft}), in which 
the temperature $T$ is effectively treated as the third collective coordinate.
The solution is evolved in small time steps by applying an explicit and unitary propagator built as a Krylov approximation of the exponential of the Hamiltonian. 
The time step is $\delta t=5\times 10^{-4}$ zs (1 zs $= 10^{-21}$ s), and the charge and mass 
distributions are calculated after $4\times10^{4}$ time steps, which correspond to 20 zs.
As in our recent calculations of Refs.~\cite{Tao2017_PRC96-024319,Zhao2019_PRC99-014618,Zhao2019_PRC99-054613,
Zhao2020_PRC101-064605,Zhao2021_PRC104-044612}, 
the parameters of the additional imaginary absorption potential that takes into account the escape 
of the collective wave packet  in the domain outside the region of calculation \cite{Regnier2018_CPC225-180} are: 
the absorption rate $r=20\times 10^{22}$ s$^{-1}$ and the width of the absorption band $w=1.0$.

The deformation collective space is divided into an inner region with a single nuclear density distribution, 
and an external region that contains the two separate fission fragments. 
The scission hyper-surface that divides the inner and external regions is determined by calculating the 
expectation value of the 
Gaussian neck operator $\displaystyle \hat{Q}_{N}=\exp[-(z-z_{N})^{2} / a_{N}^{2}]$, 
where $a_{N}=1$ fm and $z_{N}$ is the position of the neck~\cite{Younes2009_PRC80-054313}.
We define the pre-scission domain by $\langle \hat{Q}_{N} \rangle>3$, and consider the frontier of this domain as the scission surface.
The flux of the probability current 
 \begin{equation}
\label{eq:current}
J_{i}(\bm{q},T; t) = \hbar \sum_{j} \mathcal{M}_{ij}^{-1}(\bm{q},T) \rm{Im}\left( \psi^{*} \frac{\partial\psi}{\partial q_{j}} \right),
\end{equation}
through the scission hyper-surface provides a measure of the probability of observing a given pair of fragments at time $t$. 
Each infinitesimal surface element $\xi$ is associated with a given pair of fragments 
$(A_L,A_H)$ at temperature $T$, where $A_L$ and $A_H$ denote the lighter and heavier fragment, respectively. 
From the density profile obtained in the corresponding MDC-RHB calculation, 
we obtain the deformation parameters of each fragment $\beta_{2}^{L}, \beta_{3}^{L}$ and $\beta_{2}^{H}, \beta_{3}^{H}$.
By performing finite-temperature deformation-constrained RHB calculations for each fragment, 
the total binding energy at given deformations and temperature is obtained for this pair of fragments: 
$E^{L} (\beta_{2}^{L}, \beta_{3}^{L}, T)$ and $E^{H} (\beta_{2}^{H}, \beta_{3}^{H}, T)$.
From the energy balance at scission~\cite{Caamano2017_PLB770-72}, the TKE for this specific pair of fragments can be calculated as 
\begin{equation}
\label{eq:tke}
\begin{aligned}
\rm{TKE}(\xi) =& (E_{\rm B}^{\rm FS} + E^{*}_{\rm{coll}}) \\
	&-  \left[ E^{L} (\beta_{2}^{L}, \beta_{3}^{L}, T) + E^{H} (\beta_{2}^{H}, \beta_{3}^{H}, T) \right], 
\end{aligned}
\end{equation}
where $E_{\rm B}^{\rm FS}$ refers to the total binding energy of the fissioning nucleus at equilibrium minimum, and $E_{\rm coll}^{*}$ is the corresponding excitation energy of the collective initial state. 
The integrated flux $F(\xi;t)$ for a given surface element $\xi$ is defined as~\cite{Regnier2018_CPC225-180}
\begin{equation}
F(\xi;t) = \int_{t_0}^{t} dt^\prime \int_{(\bm{q},T) \in \xi} \bm{J}(\bm{q}, T; t^\prime) \cdot d\bm{S},
\label{eq:flux}
\end{equation}
where $\bm{J}(\bm{q}, T; t^\prime)$ denotes the current Eq.~(\ref{eq:current}). 
The TKE for the fission fragment with mass $A$ is defined by 
\begin{equation}
\label{eq:dtke}
\rm{TKE}(A) = \lim_{t \rightarrow \infty} \frac{\sum_{\xi \in \mathcal{A}} F(\xi;t) TKE(\xi)}{\sum_{\xi \in \mathcal{A}}  F(\xi;t)}.
\end{equation}
 The set $\mathcal{A}(\xi)$ contains all elements belonging to the scission hyper-surface such that one of the fragments has mass number $A$.

%----------------------------------------------------------------------------------------------------------------------
\section{\label{sec:results}Illustrative calculation: induced fission dynamics of $^{228}$Th}
%----------------------------------------------------------------------------------------------------------------------
In a first step, a large scale MDC-RMF calculation of $^{228}$Th is performed to generate the deformation energy surfaces, single-nucleon wave functions and occupation factors in the $(\beta_2,\beta_3,T)$ space, that determine the collective non-dissipative potential and mass tensor.  
The intervals for the collective variables are:  $-1 \le \beta_2 \le 7$ with   
a step $\Delta \beta_2 = 0.04$; $0 \le \beta_3 \le 3.5$ with a step $\Delta \beta_3 =0.05$;
and the temperature is varied in the range $0 \le T \le 2.0$ MeV, with a step $\Delta T = 0.1$ MeV.

%----
\begin{figure}[!]
\centering
 \includegraphics[width=0.45\textwidth]{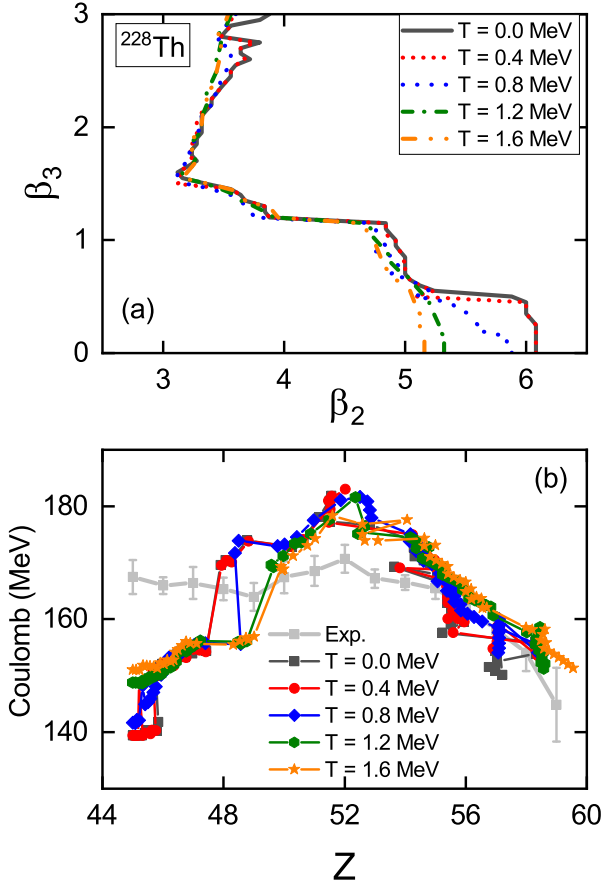}
 \caption{(Color online)~\label{fig:coulomb}%
 Scission contours of $^{228}$Th in the $(\beta_{2}, \beta_{3})$ deformation plane for several values of the nuclear temperature $T$ (a), 
 and Coulomb repulsive energies between the nascent fission fragments of $^{228}$Th, as functions of the fragment's charge, calculated at 
 different temperature $T$, compared to the experimental values of the total kinetic energy~\cite{Schmidt2000_NPA665-221} (b).
}
\end{figure}
%-----

Panel (a) in Fig.~\ref{fig:coulomb} displays the scission contours in the $(\beta_{2}, \beta_{3})$ plane for several values of the temperature $T$.
The contours generally do not differ much, especially for asymmetric fission. 
At higher temperature, the scission contour is shifted towards smaller quadrupole deformations $\beta_{2}$ values for nearly symmetric fission events.
The Coulomb repulsion for a particular pair of fission fragments can be evaluated from the relation
\begin{equation}
\label{eq:coulomb}
E_{\rm Cou} = \frac{e^2 Z_{H} Z_{L}}{d_{\rm ch}},
\end{equation}
where $e$ is the proton charge, $Z_{H}(Z_{L})$ the charge of the heavy (light) fragment, and $d_{\rm ch}$ the distance between fragment centers of charge at scission. This expression has 
typically been used to approximate the TKE of fragments in TDGCM calculations of fission dynamics \cite{Goutte2005_PRC71-024316,Tao2017_PRC96-024319,Li2022_PRC106-024307,Ren2022_PRC105-044313}. 
In Fig.~\ref{fig:coulomb} (b), we plot the distributions of Coulomb energy $E_{\rm Cou}$ at various temperatures, in comparison with the experimental TKEs obtained in photo-induced fission \cite{Schmidt2000_NPA665-221}.
One notices that although the calculated $E_{\rm Cou}$ qualitatively reproduce the trend of the data for $Z \geq 50$, 
they generally overestimate the TKEs. For $Z \leq 48$, that is, close to symmetric fission, the calculated Coulomb energies lie considerably below the experimental points. 
The values of $E_{\rm Cou}$ obtained at different temperatures are rather similar, except those at $Z \approx 48$ and near symmetric fission. The differences are obviously related to changes in the scission contours at different temperatures,  
shown in panel (a) of Fig.~\ref{fig:coulomb}. In general, irrespective of the temperature of the fissioning nucleus, 
the TKEs calculated from Eq.~(\ref{eq:coulomb}) are very similar 
since the effect of dissipation on TKEs are not taken into account.
%and the effect of dissipation on TKEs can not be taken into account. 

%----
\begin{figure}[!]
\centering
 \includegraphics[width=0.45\textwidth]{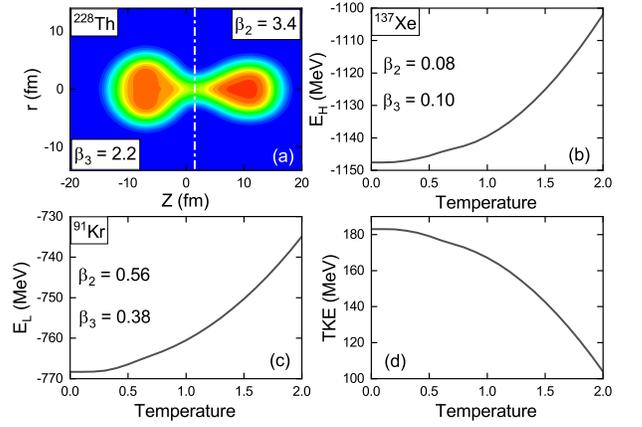}
 \caption{(Color online)~\label{fig:Z54A137}%
 Density profile of $^{228}$Th at the scission point $(\beta_{2} =3.4, \beta_{3}=2.2)$, the vertical  
 line denotes the position of the neck $z_{N}$ (a). 
 The binding energy of the heavy fragment $^{137}$Xe $(E_{L})$ as a function of temperature $T$, with the deformation
 constrained to $(\beta_{2}, \beta_{3}) \sim (0.08, 0.10)$ (b).
 The binding energy of the light fragment $^{91}$Kr $(E_{R})$ as a function of temperature $T$, with the deformation
 constrained to $(\beta_{2}, \beta_{3}) \sim (0.56, 0.38)$ (c).
The total kinetic energy as a function of temperature, calculated with the energy balance relation Eq.~(\ref{eq:tke}) (d).
}
\end{figure}
%-----

Assuming there is no evaporation of any kind before scission, the total energy of the fissioning system is stored in the nascent fragments at scission, 
as the excitation energy and kinetic energy for fully accelerated fragments $E_{\rm B}^{\rm FS} + E^{*}_{\rm{coll}} = B_{\rm eq}^{L} + B_{\rm eq}^{H} + \rm{TKE} + \rm{TXE}$,
where $B_{\rm eq}^{L}$ ($B_{\rm eq}^{H}$) refers to the total binding energy of the light (heavy) fragment at equilibrium minimum and zero temperature,  
and $\rm{TXE}$ comprises the deformation and intrinsic excitation energies deposited in the fragments at scission \cite{Caamano2017_PLB770-72}.
In Fig~\ref{fig:Z54A137} we illustrate an example of a scission point at $\beta_{2} = 3.4$ and $\beta_{3} = 2.2$. 
Panel (a) in Fig~\ref{fig:Z54A137} displays the density profile obtained from MDC-RHB calculations at zero temperature, 
the vertical line denotes the coordinate of the neck $z_N$, which separates the fissioning nucleus into two fragments 
with $Z_{H} \approx 54$, $A_{H} \approx 137$, and  $Z_{L} \approx 36$, $A_{H} \approx 91$.
The deformations of these two fragments at scission are $(\beta_{2}^{H}, \beta_{3}^{H}) \sim (0.08, 0.10)$, and $(\beta_{2}^{L}, \beta_{3}^{L}) \sim (0.56, 0.38)$.
If, just for the sake of illustration, one assumes that the deformations of these fragments do not change at scission as the temperature $T$ increases (cf. panel (a) in Fig.~\ref{fig:coulomb}), the total energies of the fragments
$E^{H} (\beta_{2}^{H}, \beta_{3}^{H}, T) = B_{\rm eq}^{H} + {\rm TXE}^{H}$ and $E^{L} (\beta_{2}^{L}, \beta_{3}^{L}, T) = B_{\rm eq}^{L} + {\rm TXE}^{L}$, 
increase quadratically with temperature $T$, as shown in Fig~\ref{fig:Z54A137} (b) and (c).
Fig~\ref{fig:Z54A137} (d) shows that the TKE calculated with the energy balance relation Eq.~(\ref{eq:tke}), decreases quadratically as $T$ increases.
We note, however, that in actual calculations discussed below, the fragment deformation at scission varies with $T$.
 
%----
\begin{figure}[!]
\centering
 \includegraphics[width=0.45\textwidth]{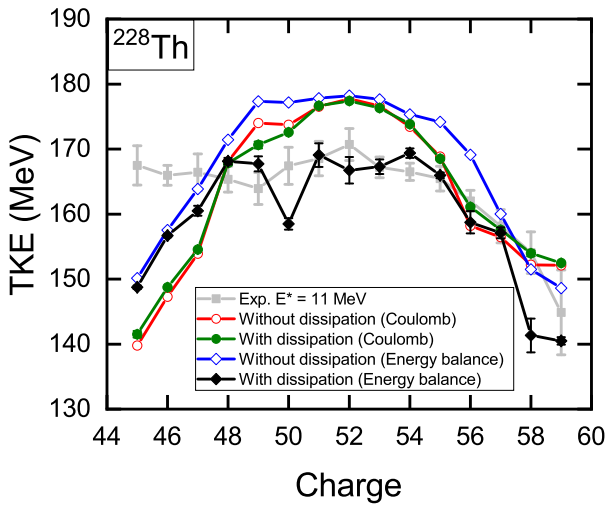}
 \caption{(Color online)~\label{fig:TKE}%
 The total kinetic energy (TKE) for induced fission of $^{228}$Th, calculated in the 3D space of axial deformation parameters $\beta_{2}$, $\beta_{3}$, and temperature $T$.
 The TKEs obtained from the energy balance formula without (blue empty diamonds), and with dissipation (black filled  diamonds), are compared with available data.
 The TKEs calculated using the Coulomb energy formula without (red empty circles) and with (green filled circles), are also shown for comparison.
 The results obtained by including the dissipation term, correspond to the mean value of four calculations with different random matrices $\eta(T, T^\prime)$.
 The resulting standard deviations are shown as error bars. 
 The data for photo-induced fission correspond to photon energies in the interval $8-14$ MeV, and a peak value of $E_{\gamma} = 11$ MeV~\cite{Schmidt2000_NPA665-221}.
}
\end{figure}
%-----

In practical 3D calculations, the two-dimensional scission contour is embedded in the three-dimensional space $(\beta_{2}, \beta_{3}, T)$.
For each scission point on this 2D scission surface, we have determined the temperature-dependent deformations $(\beta_{2}^{H}, \beta_{3}^{H})$ and $(\beta_{2}^{L}, \beta_{3}^{L})$,
from the corresponding density profiles obtained in MDC-RHB calculations, and the temperature $T$ of the fission fragments is the temperature at the corresponding scission point. 
The TKE for each pair of fission fragments at temperature $T$ (cf. panel (a) in Fig.~\ref{fig:coulomb}), is calculated using the energy balance relation Eq.~(\ref{eq:tke}).
In total, close to $2\times10^{4}$ pairs of fission fragments have been calculated.

In the next step, a full 3D calculation of induced fission dynamics of $^{228}$Th is carried out in the space $(\beta_{2}, \beta_{3}, T)$, including the dissipative coupling between deformation energy surfaces at different temperatures, as described in Ref.~\cite{Zhao2022_PRC105-054604}.
The average excitation energy of the initial state is $E^*_{\rm{coll}} = 11$ MeV. Making use of the resulting integrated flux (\ref{eq:flux}), the final TKEs are obtained from Eq.~(\ref{eq:dtke}) and plotted in Fig.~\ref{fig:TKE}. The theoretical values are compared with the data for photo-induced fission of $^{228}$Th with photon energies in the interval $8-14$ MeV, and a peak value of $E_{\gamma} = 11$ MeV \cite{Schmidt2000_NPA665-221}.
For comparison, in Fig.~\ref{fig:TKE} we have also included the resuls obtained using Eq.~(\ref{eq:dtke}), but with ${\rm TKE}(\xi)$ calculated with the Coulomb energy expression Eq.~(\ref{eq:coulomb}) (cf. panel (b) in Fig.~\ref{fig:coulomb}).
Theoretical values obtained without dissipation are denoted by empty symbols, while those for which the integrated flux is obtained by including the dissipation term with the function $\bm{\eta}(\bm{q}; T, T^{\prime})$ in Eq.~(\ref{eq:tdgcmgoa_ft}),  
are denoted by black filled diamonds (energy balance formula), and green filled circles (Coulomb energy).

Without dissipation, the TKEs obtained using either the energy balance formula Eq.~(\ref{eq:tke}), or the Coulomb energy  Eq.~(\ref{eq:coulomb}), generally overestimate the experimental values, except for the region of symmetric fission with $Z \leq 47$. The TKE values calculated using the energy balance are somewhat larger than the ones from the Coulomb repulsion, because the former include the prescission collective kinetic energy that is missing in the Coulomb energy expression.

When the dissipation term is included in the Hamiltonian of Eq.~(\ref{eq:tdgcmgoa_ft}) for the time evolution of the statistical wave function, the resulting TKEs are denoted by black filled diamonds (energy balance), and green filled circles  (Coulomb repulsion) in Fig. ~\ref{fig:TKE}. Since the matrix elements of the dissipation function $\bm{\eta}(\bm{q}; T, T^{\prime})$ Eq.~(\ref{eq:eta}) are assumed to be 
Gaussian random variables, the calculation has been carried out with four different random matrices $\bm{\eta}(T,T^{\prime})$. We plot the mean values, and the corresponding standard deviations are shown as error bars. 
The strength of the dissipation term is determined by the parameter $\gamma = 0.01$, and this is the same value used in Ref.~\cite{Zhao2022_PRC105-054604} to calculate the charge yields of $^{228}$Th.
The results obtained using the Coulomb energy in Eq.~(\ref{eq:dtke}), are almost identical to those obtained without dissipation (the error bars are too small to be visible). 
This is because the Coulomb repulsion does not depend on the temperature of the fragments and, therefore, by equating the TKE with Coulomb energy between the fragments at scission, the dynamic effect of dissipation cannot be taken into account. 
The small differences arise because the scission contour depends on temperature, as shown in panel (b) of Fig.~\ref{fig:coulomb}. 
The total kinetic energies obtained when ${\rm TKE}(\xi)$ in Eq.~(\ref{eq:dtke}) is calculated using the energy balance formula Eq.~(\ref{eq:tke}), are shown as black filled diamonds in Fig. ~\ref{fig:TKE}. The inclusion of dissipation generally reduces the calculated TKEs, bringing them in quantitative agreement with the data. This is because 
dissipation heats up the fissioning system and, with the fragments at higher temperature, there is less energy available as kinetic energy. This is also clearly seen in Fig. 7 of Ref.~\cite{Zhao2022_PRC105-054604}, where the time-integrated collective flux through the scission contour is shown as a function of temperature, without and with the dissipation term included. It was shown that dissipation broadens the distribution of the flux and extends the high-T tail to higher temperatures. In this particular example, however, we note that dissipation appears to have very little effect on the calculated TKEs in the region of symmetric fission with $Z \leq 47$.

\section{\label{sec:summary}Summary}

 Starting from an extension of the TDGCM that includes dissipation in the description of induced fission dynamics \cite{Zhao2022_PRC105-054604}, we have developed a method to calculate the corresponding distribution of total kinetic energies as a function of charge or mass of the fission fragments. In ordinary applications of the TDGCM, the total kinetic energy for a particular pair of fragments is evaluated as the energy of their Coulomb repulsion at scission. Standard TDGCM by definition describes non-dissipative dynamics and, in the adiabatic approximation, all the potential energy is converted into collective kinetic energy during the saddle-to-scission evolution. As a result, the calculated TKEs generally overestimate the experimental values \cite{Ren2022_PRC105-044313}. 
 
In Ref.~\cite{Zhao2022_PRC105-054604} we have derived an approximate expression for a dissipative term in TDGCM, that propagates a statistical collective wave function through deformation energy surfaces at different temperatures and, therefore, can be used to model the heating of the fissioning system. To determine the corresponding TKE distributions, fully self-consistent mean-field calculations have to be performed in the 3D space of collective coordinates - quadrupole and octupole deformations, and nuclear temperature. 
At each temperature and at each point on the scission contour, the TKE for the corresponding pair of fragments is calculated using the energy balance at scission. This requires evaluating the energies of the deformed fragments at scission, for each value of the temperature. To calculate the TKE distribution, the kinetic energies of the fragments at a given temperature are folded with the time-integrated flux of the probability current (cf. Eq.~(\ref{eq:dtke})). 

The method has been illustrated with an example of the TKE distribution for induced fission of $^{228}$Th. It is shown that, without the inclusion of the dissipation term, the TKEs calculated using energy balance at scission are very similar to those obtained from the Coulomb repulsion between the fragments at scission and, therefore, the theoretical values overestimate the data from photo-induced fission \cite{Schmidt2000_NPA665-221}. By including the dissipation term in the Schr{\"o}dinger equation for the time-evolution of statistical collective wave functions, we are able to describe the heating of the fissioning system and this reduces the collective kinetic energy at scission. With the strength parameter of the dissipation term adjusted to reproduce the fission charge yields \cite{Zhao2022_PRC105-054604}, the resulting TKEs are in quantitative agreement with the experimental values. We note that, although this extension of the TDGCM to include a temperature-dependent dissipation mechanism enables calculation of both the charge (mass) yields and TKEs in quantitative agreement with data, the method is computationally very intensive and computer-time consuming. Nevertheless, it is worth performing additional tests, for different fissioning systems, and different excitation energies. In addition, as already noted in the first part of this work \cite{Zhao2022_PRC105-054604}, a fully microscopic form of the dissipation term, that would also completely determine its strength thus eliminating the need for an adjustable parameter, necessitates  an extension of the GCM to generating states that are functions not only of collective coordinates, but also of collective momenta conjugate to $\bm{q}$ (dynamical GCM). 
%%%%%%%%%%%%%%%%%%%%%%%%%%%%%%%%%%%%%%%%%%
\bigskip
%---------------------------------------------------------
\acknowledgements
This work has been supported in part by the QuantiXLie Centre of Excellence, a project co-financed by the Croatian Government and European Union through the European Regional Development Fund - the Competitiveness and Cohesion Operational Programme (KK.01.1.1.01.0004) and the Croatian Science Foundation under the project Uncertainty quantification
within the nuclear energy density framework (IP-2018-01-5987).
It has also been supported by the National Natural Science Foundation of China under Grant No. 12005107 and No. 11790325.
The authors acknowledge the Beijing Super Cloud Computing Center (BSCC) for providing HPC resources 
that have contributed to the research results reported within this paper \url{URL: http://www.blsc.cn/}.

%\bibliographystyle{apsrev4-1}
%\bibliography{nuclear,Nuclear-Fission,MyOwn}

\end{document}